# ERIC: An Efficient and Practical Software Obfuscation Framework

Alperen Bolat[a], Seyyid Hikmet Çelik[a], Ataberk Olgun[a], Oğuz Ergin[a], Marco Ottavi[b]
[a]TOBB University of Economics and Technology, Turkey,
[b]University of Twente, The Netherlands and University of Rome Tor Vergata, Italy
[a]{alperenbolat, seyyidhikmetcelik, aolgun, oergin}@etu.edu.tr, [b]m.ottavi@utwente.nl

*Abstract*—Modern cloud computing systems distribute software executables over a network to keep the software sources, which are typically compiled in a security-critical cluster, secret. However, these executables are still vulnerable to reverse engineering techniques that can extract secret information from programs (e.g., an algorithm, cryptographic keys), violating the IP rights and potentially exposing the trade secrets of the software developer. Malicious parties can (i) statically analyze the disassembly of the executable (static analysis) or (ii) dynamically analyze the software by executing it on a controlled device and observe performance counter values or exploit side-channels to reverse engineer software (dynamic analysis).

We develop ERIC, a new, efficient, and general software obfuscation framework. ERIC protects software against (i) static analysis, by making only an encrypted version of software executables available to the human eye, no matter how the software is distributed, and (ii) dynamic analysis, by guaranteeing that an encrypted executable can only be correctly decrypted and executed by a single authenticated device. ERIC comprises key hardware and software components to provide efficient software obfuscation support: (i) a hardware decryption engine (HDE) enables efficient decryption of encrypted hardware in the target device, (ii) the compiler can seamlessly encrypt software executables given only a unique device identifier. Both the hardware and software components are ISA-independent, making ERIC general. The key idea of ERIC is to use physical unclonable functions (PUFs), unique device identifiers, as secret keys in encrypting software executables. Malicious parties that cannot access the PUF in the target device cannot perform static or dynamic analyses on the encrypted binary.

We develop ERIC's prototype on an FPGA to evaluate it end-to-end. Our prototype extends RISC-V Rocket Chip with the hardware decryption engine (HDE) to minimize the overheads of software decryption. We augment the custom LLVM-based compiler to enable partial/full encryption of RISC-V executables. The HDE incurs minor FPGA resource overheads, it requires 2.63% more LUTs and 3.83% more flip-flops compared to the Rocket Chip baseline. LLVM-based software encryption increases compile time by 15.22% and the executable size by 1.59%. ERIC is publicly available and can be downloaded from https://github.com/kasirgalabs/ERIC.

*Index Terms*—Software Obfuscation, Trusted Execution, Hardware Authentication

## I. INTRODUCTION

With the rising trend in the use of electronic devices in the context of the Internet of Things (IoT), more and more devices communicate with each other over networks [33], [20], [22], [1]. Maintaining secure and reliable communication channels between multiple devices over a network is important to provide security and reliability guarantees for a wide variety of applications [3], [53].

Embedded systems typically consist of processors or processing units that execute a program. These systems often serve a specific purpose and are not as capable as general-purpose systems in terms of performance, power budget, and computational capability [35], [18]. Programs that are executed by embedded systems are often compiled in more capable computing systems (e.g., personal computers, cloud systems) and their compiled binaries are delivered to the embedded systems over physical interfaces (e.g., serial connection) or over a network [21], [6], [26], [21]. Typically, these binaries implicitly encapsulate critical information (e.g., architectural details of the target hardware and sophisticated, trade secret algorithm implementations) [38], [50]. It is important that this information is not exposed to malicious parties.

We identify two types of attacks that threaten the secrecy of critical information embedded in binaries. First, a binary can be converted into a human-readable form by using standard compiler tools (e.g., disassembler) and can be analyzed to identify critical information [15], [27], [45], [5], [51]. We refer to these types of attacks as static-analysis attacks. Second, a binary can be executed on a computer that is controlled by malicious parties and the computer's state (e.g., performance counters, register values) can be monitored to reverse engineer the source code from the binary. We refer to these types of attacks as dynamic-analysis attacks [44], [46], [42], [4], [10]. Our goal is to design a new framework that can prevent both static and dynamic-analysis attacks.

We propose ERIC, a new framework that keeps program information secret regardless of how the program is transferred from its source to the destination hardware platform. To do so, ERIC uses cryptographic keys that are generated from an identifier unique to a target hardware device (e.g., physical unclonable functions) to encrypt program binaries. The encrypted program can only be decrypted by the target hardware device, preventing malicious parties from performing static and dynamic analyses.

ERIC is composed of two components: First, a new compiler supports partial and complete encryption of software executables (i.e., program binary). Second, a hardware decryption engine enables efficient decryption of encrypted binaries. ERIC uses physical unclonable functions (PUFs) to generate cryptographic keys that are used in symmetric encryption and decryption of program binaries. In this way, ERIC guarantees that an encrypted binary can only be decrypted and executed by the target hardware.

We develop ERIC's FPGA-based prototype and integrate it into a real RISC-V system to evaluate ERIC end-to-end. We implement ERIC's compiler by augmenting LLVM to perform encryption on RISC-V binaries. Our prototype builds on open source projects and itself is open sourced on

https://github.com/kasirgalabs/ERIC. We hope that our open source prototype will be of use to industry and researchers going forward.

We evaluate ERIC end-to-end using multiple workloads to understand how its hardware and software components perform and affect the overall performance of the system. Our results show that compilation time increases by 15.22% on average compared to the baseline compiler without encryption and end-to-end execution time increases by no more than 7.05% across all workloads. ERIC's hardware components introduce a minor 3.17% FPGA resource overhead and can be implemented at the low FPGA component cost.

ERIC makes the following contributions:
- ERIC is the first end-to-end software obfuscation and trusted execution framework that offers a lightweight architecture applicable to all systems.
- We implement and publicly released an end-to-end prototype of ERIC which comprises a real RISC-V system and a compiler.

## II. BACKGROUND AND THREAT MODEL

### A. LLVM

LLVM is a set of compiler and toolchain technologies that can be used to develop any programming language as a front-end and any instruction set architecture as a back-end [28], [31]. LLVM is designed around a language-independent methodology called Intermediate Representation (IR). The program compiled using any language is first translated to the IR language. IR is a high-level assembly language that can be optimized with a variety of transformations over multiple passes. Many optimizations and analyses are possible during compilation with IR. After all the desired optimizations and analyses are made on the IR representation of the program, a binary code translation for the target instruction set architecture (ISA) is produced from the IR language.

With the IR methodology, it is possible to develop a custom compiler on LLVM. The designed compiler can be built by performing the target compilation options on the IR language. In addition, this compiler is compatible with most languages thanks to the support of LLVM libraries.

### B. Arbiter PUF

Physical unclonable function (PUF) is a function that provides a physically defined "digital fingerprint" output (response) for a given input and conditions (difficulty), usually serving as a unique identifier for a semiconductor device [34], [24]. PUFs are mostly based on variations that occur naturally during semiconductor fabrication. Since the distribution of these variances is unique for each semiconductor, PUFs use it to obtain the unique identity of the device. There are many PUF methods proposed to date [8], [13], [37], [40], [7], [25], [19], [14]. Delay-based PUFs are among the most common PUF methods used [36].

Arbiter PUFs compare the delay of two identical paths to generate a '0' or '1' bit, depending on the result of the comparison [32]. Although no two paths are the same and should introduce the same delay, minor unforeseen differences during the manufacturing process make one path ultimately faster than the other.

Arbiter PUFs use challenge–response authentication. Challenge-response authentication is the protocol in which the source gives a response output for an incoming challenge input. PUF-based systems can check system authentication by the validation of the response output. Figure 1 shows the scheme of the 5-bit challenge and 1-bit response PUF model. As seen in the figure, the response output changes depending on the path of the delays on the hardware and the challenge value.

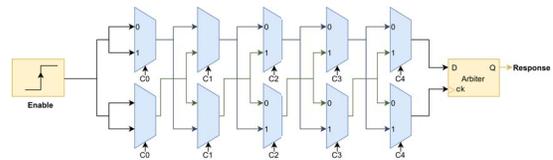

Fig. 1: 5 Bit Challenge - 1 Bit Response PUF Scheme

### C. Threat Model

In our threat model, we assume that the executable (program binaries) is transmitted over an untrusted network. Malicious parties can retrieve the executable to violate IP rights, make modifications to the executable and send the modified version to the destination hardware. We assume that the target hardware is trusted. In summary, we protect against these types of threats: (i) Hijacking critical programs and resulting reverse engineering applications, (ii) attempting to execute programs of unknown origin on user hardware, (iii) running programs compiled by the software source on unlicensed or unverified hardware and (iv) the execution of malicious modifications or soft errors to the program on the system.

## III. ARCHITECTURE OF ERIC

ERIC is an efficient and practical framework designed to establish a trusted execution environment that provides authenticated target-hardware-specific software compilation capability. To provide such capability, first, ERIC encrypts software such that the encrypted software can only be decrypted in the target hardware, protecting the software against malicious attacks in the form of static and dynamic analyses, second, ERIC integrates a secure hash digest of the software within the software's encrypted version to support integrity validation. ERIC consists of two components: (i) hardware-based architecture and (ii) software-based architecture.

ERIC's hardware and software components are easily decoupled. For example, ERIC's software components can be used in one computer to compile and encrypt programs, and its hardware components can be used in another computer to perform decryption. This way, an encrypted program can be securely transferred over a network from one computer to another, and its integrity can be validated by the hardware that decrypts and runs it. If the integrity of an encrypted program can be validated, it is guaranteed to come from a trusted source, thus the encrypted program's authenticity is also validated. This way, the program runs only on the target hardware and the target hardware only executes the programs written for it. We refer to this feature of ERIC as *two-way authentication*. Figure 2 shows how two-way authentication works between hardware and software interfaces.

On the other hand, ERIC achieves obfuscation of the program binaries from malicious activities while transferring between the software source and the target hardware.

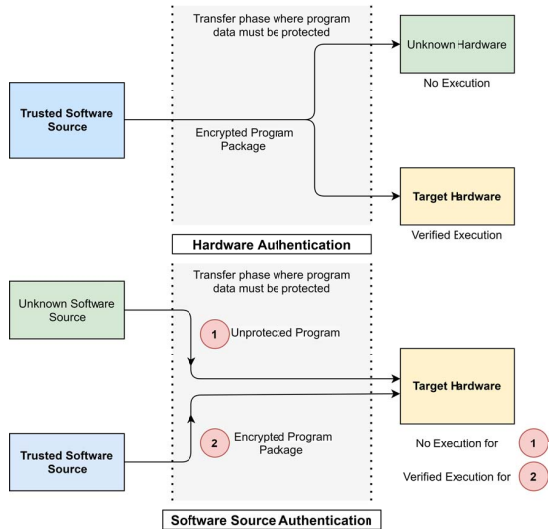

Fig. 2: Two-Way Authentication Model

*1) Software Architecture*

Target-hardware-specific encryption is provided by a PUF-based key. For this, the compiler performs integrated encryption with the PUF-based Key Generator Unit on the hardware side. PUF-based keys are obtained by passing the PUF key in the hardware through the function (e.g., secure hash algorithm [16]) in the Key Management Unit. Since the PUF key is immutable on each target hardware, an encryption mechanism directly linked to the PUF key gives unlimited access to the trusted program resource, and this access relationship cannot be changed. To avoid this, it is suggested to obtain a PUF-based key from the PUF key that the target hardware can configure at any time. The use of PUF keys provides a unique key for each device that has physical foundations and does not need to be constantly stored in a register.

As stated in Section II.C, the proposed architecture aims at the obfuscation and safety of the compiled program until the execution stage while achieving two-sided authentication between the target hardware and the software source. Therefore, it is assumed that the handshake is already done for the hardware targeted by the software source, and PUF-based keys that are compatible with the target hardware are assumed to be known to the software source. The key unit we added on the compiler side takes the PUF-based key of the target hardware as input and generates the keys to be used in encryption operations. The PUF key is not exposed to the user during software compilation. With the abstraction layer created between the PUF-based key generator in the hardware and the key generator in the software, there is no need for direct access to the PUF key when integrated encryption with the PUF key. This allows long-term key usage, enabling different key configurations in the system. Also, in this way, the PUF key in the hardware is protected and can be used for different targets since it is not shared with the software developer directly.

ERIC's software components encrypt the program in various configurations during the compilation phase. The encrypted program is intended to remain secure until it reaches the target hardware. The main goals of the encryption are to be protected from reverse engineering methods, malicious add-ons that can be applied to the program or bit flips that may occur because of soft errors while the program is being transferred or stored. ERIC includes a graphical interface in order to understand the encryption stages more easily and to enable the programmer to control the encryption according to their own needs. Encryption can be done through this interface according to the requirements.

The proposed architecture is compatible with different encryption methods. New encryption algorithms can be easily implemented in the system by the ERIC's interface. With the changes to be made in the target files, the user has the freedom to upload his own encryption method to the system. After deciding on the encryption algorithm, encryption methods should be determined. There are three different encryption methods that can be used for this. These are the complete encryption of the program, partial encryption of the program, and the partial encryption of a select few instructions of the program by specifying the target bits in the instruction encoding.

All instructions are encrypted with PUF-based keys and ready to be executed on the target hardware. For the target hardware to detect which instructions are encrypted, the encryption map must be transmitted to the other party along with the encrypted program. Using partial encryption, the programmer can protect the critical parts of the program or create an area within the program that can only be active on the target hardware. In addition, the programmer can select the features he/she wants to run only on licensed hardware within the program.

The interface where target instructions can be selected for partial encryption is provided with ERIC. The presented interface also allows selecting special parts within the target instructions. In this way, only critical information can be protected without interfering with the program flow. For example, only the pointer values of the instructions that make memory accesses can be encrypted, which makes it difficult to follow the program's memory trace. If the opcode parts of the instructions are not encrypted during partial encryption, it will also make it difficult to understand that the program is encrypted in the case of reverse engineering.

ERIC is suitable for compiling from a single software source for multiple target hardware or creating multiple trusted software sources for single target hardware. For this, it is only necessary to apply the appropriate conversion function to the system to match the PUF-based key with the PUF keys. Also, if the hardware manufacturer maps two or more different hardware to the same PUF-based key while performing the conversion function in the Key Management Unit, programs can be created to run on multiple hardware of their own with a single compile step. This implies that ERIC does not have a scaling problem for multiple targets or sources.

In addition to encryption, ERIC generates signatures for compiled programs to ensure that the program reaches the target without modification. This signature is obtained by running a cryptographic hash function on the instructions before the program is encrypted. The signature is produced before the program is encrypted and the signature is encrypted with the program, making the signature useless for those who cannot decrypt the program. The resulting signature is finally encrypted with a PUF-based key. In this way, interference or errors that may occur in the program during the transfer or storage of the program become detectable by the hardware.

The developments made on the software side of ERIC are completely suitable for targeted customization. The variety of encryption methods, the ability to change the encryption function, and the selection of ISA specifications suitable for the target hardware make this possible.

*2) Hardware Architecture*

For the hardware to run the encrypted and signed program securely and efficiently, hardware architecture is needed that will work in harmony with the software architecture. ERIC's Hardware architecture includes several units. These are the Decryption Unit, the Validation Unit, the Key Management Unit, the Signature Generator, and the PUF Key Generator. The high-level unit in which these units are integrated is called as Hardware Decryption Engine (HDE) Unit. The hardware architecture is completed by integrating the processor with the HDE Unit. The security model in the hardware ensures that the received programs are kept encrypted until they are loaded into the main memory for execution. Since decryption operations are performed without writing the program to memory, the recommended hardware architecture is compatible with common processor architectures.

**PUF Key Generator (PKG).** The PKG enables the generation of keys that act as an identity for the hardware device due to the differences in the hardware during production (e.g., process variation). These keys provide the hardware with a unique identification number. These PUF keys will be used in ERIC to distinguish between two different hardware.

**Key Management Unit.** PUF-based keys used by hardware are used to decrypt the incoming encrypted program. In order to integrate the keys used on the software side and the keys used on the hardware side, the existing PUF key goes through the key generation function within the Key Management Unit. The PUF-based key is obtained by passing the PUF key through the function. In this way, multiple PUF-based keys are generated with a single PUF key. As mentioned earlier, this provides a layer of abstraction between the encryption/decryption key and the PUF key. Using the PUF key with abstraction allows configuring the function used for the PUF-based key in the Key Management Unit, allowing to change the compatible software resources according to time or preferences. With this flexibility, if the necessary variables in the hardware are given as input to the PUF-based key generation function a program that can only be decrypted and run at a specific time range or a program that can only be decrypted at a specific temperature, frequency, or altitude, etc. can be obtained. We did not discuss the different configurations of PUF-based key generation for the sake of simplicity of the paper and planned for future work.

**Decryption Unit.** The Decryption Unit decrypts the program that reaches the SoC encrypted with the PUF-based key. If the program is partially encrypted, the Decryption Unit analyzes the additional bits added to the encrypted program for each instruction. While these are the addition of a new bit to each instruction in instruction-based partial encryption, the use of bits indicating the encryption to be used may increase depending on the selection of partial encryption. The function to be applied for decryption need to be the reverse pair of the encryption function on the compile process. In this way, it is ensured that the same data is obtained again.

**Signature Generator.** Incoming programs to the SoC are carried with signatures obtained before program encryption in the software architecture. These signatures are also encrypted with the program. The Signature Unit is used to recalculate the signature from the decrypted program in the hardware. This unit creates the signature of the program with the instructions it receives as the program is decrypted. Finally, the generated signature is transferred to the Validation Unit.

**Validation Unit.** When the decryption of the program is finished, it transfers its own calculated signature to the Validation Unit. Likewise, the encrypted signature that comes with the program is transferred to the Validation Unit. After decrypting the signature that comes with the program in the Validation Unit, the signature calculated by the hardware itself and the signature that comes with the program are compared. If there is a match, the decrypted program is authorized for execution.

Figure 3 shows the schematic of the proposed architecture. The orange color on this figure represents the key components of ERIC. The numbers on the figure describe a typical workflow of ERIC from program encryption to execution on the SoC.

❶ refers to the generation of PUF-based keys in hardware, which is the first step required for the architecture to work. The PUF key obtained with the PUF Key Generator generates the PUF-based key, which will be required for encryption and decryption, with the Key Management Unit. In this way, it also provides an abstraction layer over the PUF Key, which is critical data for SoC's security.

❷ step is to give the requirements to the compilation stage so that the correct encryption can be done. ERIC's graphical interface and technical documentation can be used for this step. The ISA targeted by the program, the function to be used for encryption, the partial or full encryption decision, and the key information of the target hardware should be determined.

❸ step involves encrypting the program and packaging it with the signature. First, the program is compiled for the target ISA using compiler libraries according to the requirements determined in the previous step. After the compilation is finished, the signature of the program is obtained with the Signature Generator. Second, the key management function, using the PUF-based key transferred to the compiler stage, generates keys suitable for the encryption function. The obtained signature and keys are moved to the encryption function with the compiled program. In this function, the program is encrypted according to the encryption constraints of the previous step and, if necessary, packed with the extra information needed to decrypt it (in case ERIC performs partial encryption of a program). Then, with the encryption of the signature, the encrypted program package and the signature are ready to exit from the software source.

❹ step refers to the secure packaging of the program until it reaches the target hardware. The program may be stored on a server waiting to be requested by the target hardware, or it may be compiled as one of the sub-threads in distributed systems. It may also be sent to remote hardware from a program source. In this situation, if the program is accessed by non-target hardware or malicious parties, the program is protected by the encrypted program package and the encrypted signature.

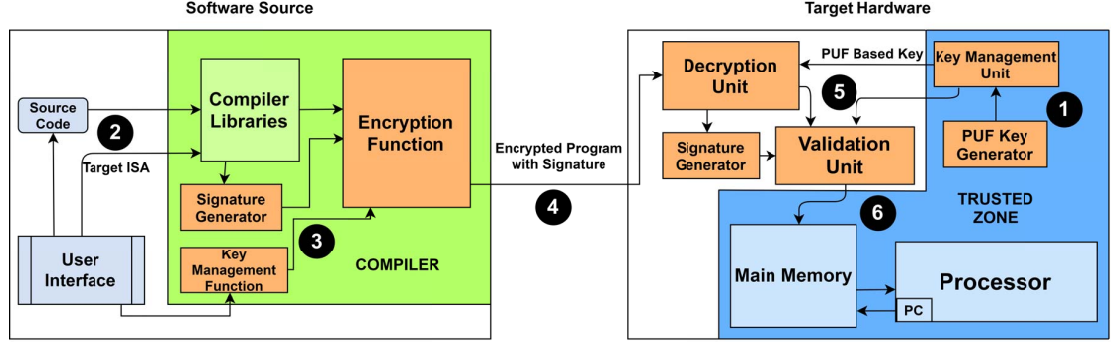

Fig. 3: ERIC's Architecture

**5** At this stage, the program and its signature that reaches the hardware are decrypted in the Decryption Unit with the PUF Based Key. The decryption process is done in accordance with the configurations determined in the encryption process. Then, the decrypted program is used to generate signatures again in the Signature Generator Unit. In addition, the decrypted signature is transferred to the Validation Unit.

**6** step is the final stage of the flow. In the Validation Unit, the signature that comes with the program is compared with the signature recalculated in the Signature Generator of hardware. In the case of a match, it is understood that the program has come without any changes and is specially packaged for this hardware, and the decrypted program is sent to the Trusted Zone and becomes suitable for executing on the processor.

## IV. IMPLEMENTATION AND TEST RESULTS

In order to quantify the performance overhead introduced by ERIC, we implement ERIC on (i) software source and (ii) target hardware.

TABLE I: Test Environment

| Parameter | Value |
| --- | --- |
| **FPGA** | Xilinx Zedboard [52] |
| **PUF Type** | Arbiter PUF |
| **PUF Parameters** | 32x 8-bit challenge 1-bit response |
| **Signature Function** | SHA-256 |
| **Encryption Function** | XOR Cipher |
| **SoC** | Rocket Chip (In-Order 6-stage) [9] |
| **Test Frequency** | 25 MHz |
| **Target ISA** | RV64GC |
| **L1 Data Cache** | 16KiB, 4-way, Set-associative |
| **L1 Instruction Cache** | 16KiB, 4-way, Set-associative |
| **Register File** | 31 Entries, 64-bit |

We perform the evaluation of ERIC in two parts. First, we measure ERIC's encrypted compilation performance for various software sources. Second, we evaluate ERIC's performance in decrypting the encrypted software binary on the target hardware. Table I shows the configuration of our evaluation. We tested the software source that is encrypted for the target hardware on our custom LLVM-based encryption compiler design. We implement the target hardware by building the SoC we designed for the proposed architecture within the FPGA. Figure 4 shows the schematic of these designs. MiBench [23] is used as a benchmark when evaluating system performance. We selected benchmark programs of MiBench which is capable with LLVM and RISC-V architecture. Since the framework we proposed is based on iterations on the program and is directly related to the program size in memory, it is also aimed to use programs of different sizes as benchmarks.

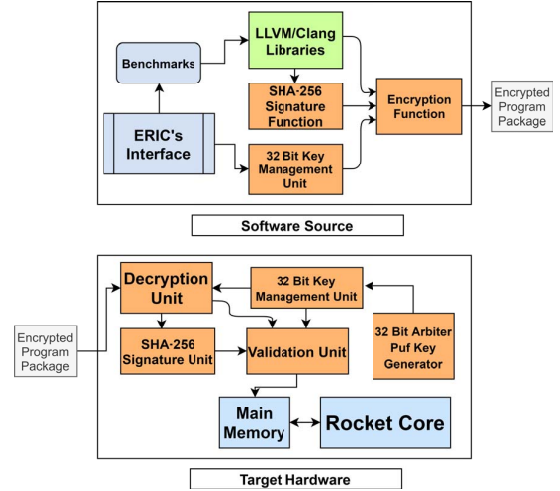

Fig. 4: Scheme of Implemented Model

### A. Software Source

We evaluate the software source in a system that includes the custom compiler we designed. We run benchmark programs on this system, in which we integrate the encryption and signature generation functions with our custom compiler. The compiler is ported from the Clang compiler driver. Also, encryption and signature generation mechanisms are ported from LLVM tools for integration with the compiler. LLVM provides functions that support instruction sets and their extensions, so instructions can be selected directly according to their flags during compilation. Also with LLVM support, ERIC allows optimizing the distribution of encrypted instructions during compilation in accordance with preferences. The mechanism that we build is combined with ERIC's interface where compilation and encryption configurations can be selected

We use LLVM 11.1 and Clang 11.1 versions for our prototype. To create the signature mechanism, we implement the SHA-256 function in C++ and the XOR cipher function for the encryption function. Since the XOR cipher function is an encryption method made by passing instructions through successive XOR gates, the encrypted message is accessed back in symmetrical steps. We use this function for the simplicity of the design. The encryption function is performed with the keys from Key Management Unit. The Key Management Unit converts the PUF-based key given from the user interface into keys in the appropriate formats for the Encryption Unit. In this way, multiple encryption iterations continue with a single PUF-based key. A PUF-based key is a hardware-generated key using a PUF key and it is implemented by the Key Management Unit on hardware. How we implemented this process is explained in Section IV.B.

To demonstrate the performance of a system that compiles programs by encryption and generates signatures, we show the results with variation in compile-time and program size.

First, the constant change in program size is the addition of signatures. Regardless of its size, every program is packaged by adding a 256-bit signature due to the SHA-256 algorithm. The dynamic increase in program size does not exist if all instructions are encrypted. However, when the program is partially encrypted, a bit is added for each instruction in the program, indicating whether the instruction is encrypted or not. For partial encryption configuration, the instructions randomly determined are selected for encryption from the program. This means a 1-bit increase in program size for every instruction in the program. As a result, if the program is fully encrypted, only a 256-bit signature increase will be seen. On the other hand, if the program is partially encrypted based on instruction selection, the program package size will increase by 1 bit for each instruction and by 256-bits for the signature. Program package size converging to this calculation is measured in tests. It has been observed that the rate of increases in program package size can change since 1 bit of extra information is received for 16 bits if the compressed instructions in the RISC-V ISA are included in the program.

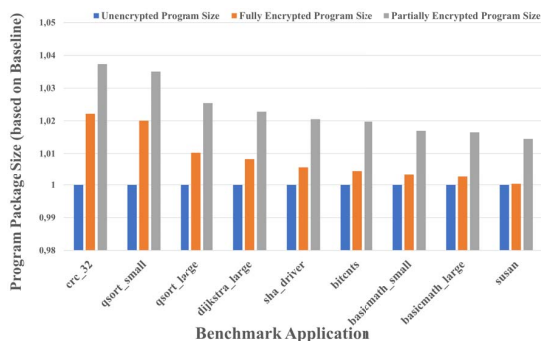

Fig. 5: Program Package Size Comparison of Encrypted Program Packages and Unencrypted Compiled Programs based on its Unencrypted Program's Size Normalization

Since fixed-size signature bits are added to each program package regardless of the size of the program, it is expected that the program size change rates will not be equal. Figure 5 shows the size change of encrypted program packages relative to plain-text (i.e., not encrypted) program size. The highest increase, the program size packed with the signature in encrypted form, is 3.73% more than the normal compiled program size and the average increase is 1.59%.

The change in compile time of Benchmark Programs is shown in Figure 6. Each benchmark program is normalized to its baseline and shown in the graph. To obtain the baseline, each program was compiled with the default Clang compiler and the compilation time was measured. Then, the time taken by compiling and packaging the programs was measured with the mechanism we implemented in the same environment. According to the results obtained, the compilation time increased by 33.20% in the worst scenario and 15.22% on average.

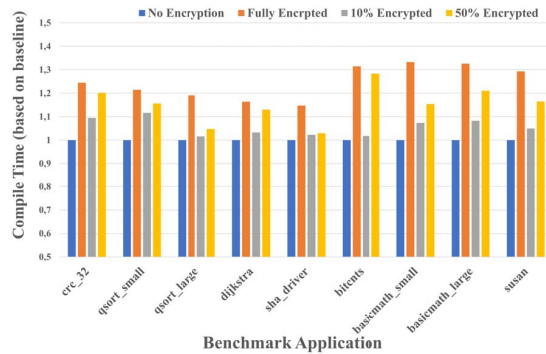

Fig. 6: Compile-Time Comparison of Each Benchmark Application based on its Unencrypted Program's Compile-Time Normalization

## B. Target Hardware

We evaluate the performance of the target hardware by implementing the HDE Units, which we develop as the mechanism that provides security with Rocket Chip, on Xilinx Zedboard. XOR Cipher-based decryption is used for the Decryption Unit. By connecting these units with a common interface, the area overhead in the hardware was calculated. FPGA implementation of Rocket Chip with the HDE Unit and only Rocket Chip implementation are compared in Table II. The proposed hardware architecture requires 2.63% more lookup tables (LUTs) and 3.83% more flip-flops compared to the Rocket Chip baseline.

TABLE II: Area Results of FPGA Implementation

|  | Rocket Chip | Rocket Chip + HDE | Change (%) |
|---|---|---|---|
| Total Slice LUTs | 33894 | 34811 | +2,63 |
| Total Flip-Flops | 19093 | 19854 | +3,83 |
| Frequency(MHz) | 25 | 25 | - |

We ran the encrypted program packages on the FPGA to observe the performance of the encrypted programs on the SoC. In order to create a baseline, we ran the programs compiled without encryption in the same system configurations with Rocket Chip. The change in execution time is shown in the graph in Figure 7, normalized to the baseline. According to the results obtained, it is observed that the method we recommended slows down the system by 7.05% at most and 4.13% on average. Since the architecture proposed by ERIC is outside of the Rocket Chip, the effect of the working performance and working methods of the programs on the system performance is not directly observed when working with ERIC. However, there is a direct proportionality between the dynamic size of the program and the performance.

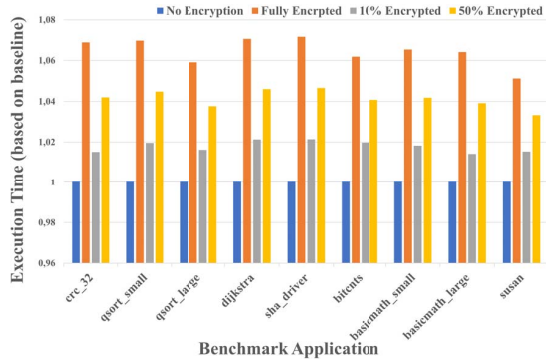

Fig. 7: Execution Time Comparison of Each Benchmark Application based on its Unencrypted Program's Execution

## V. Related Work

To the best of our knowledge, ERIC is the first software obfuscation and trusted execution framework that is fully end-to-end, from the encrypted compilation on software to trusted execution on the hardware. ERIC enables programs to be securely packaged as they send the compiler and only execute on target hardware without any change to the original program. ERIC's architecture and implementation are described in sections III and IV.

In this section, we briefly describe other related works about software obfuscation and trusted execution. In [30], [29], for ensuring memory security and program authentication, the entire memory is encrypted with the message authentication code. In these studies, every line in the memory is protected with AES encryption. The inadequacy of this study is that it uses AES encryption for each line in the memory and cannot create a secure area for the main memory. There is the high memory latency that AES brings. Programs with poor cache performance experience an extra delay each time when trying to access the main memory. In [47], [49], [48], the encryption-based mechanism is proposed to prevent attacks based on hardware and software. It performs PUF-based encryption and uses a security tree consisting of different levels in memory. They offer a toolchain for developing secure software for their architecture that includes a secure operating system to manage different levels of memory protection. However, the proposed architecture is built for a specific SoC and cannot be compatible with all processors or ISAs. The applicability of the system becomes difficult due to changes in the operating system. In addition, there is an average of 30% slowdown in instruction-per-cycle performance relative to the baseline. Another study provides a security layer with an isolated instruction set extension in hardware. In [39], encryption is made by PUF keys. Security of the memory and the program is provided by encryption with this key. It supports systems built according to the protocol between the hardware manufacturer and the software manufacturer. The change in the instruction architecture causes the general compilation tools to become unavailable. This study, unlike our study, does not encrypt the program. Instead, caches have an architecture where they work like main memory, and work is focused on translation lookaside buffer (TLB) performance. In contrast, the architecture we proposed is applicable to all processors, as it decrypts the instructions without fetching them to the processor core. It does not directly affect the execution process and can be configured for all compilers as it supports standard ISAs.

On the other hand, there are proposed trusted execution environment studies [11], [41]. Mainly, these studies are approaches to isolating execution within the processing unit. However, the goal of ERIC is to protect the program from malicious actions and any modification during transit in an untrusted network medium while ensuring that the program comes from the correct software source and reaches the correct target hardware. Also, ERIC does not directly involve the architecture of the processing unit.

There are also prior works for the security of the program on the processor and the execution of the trusted program [54], [12], [2], [17], [43]. ERIC framework stands out with its features: (i) provides a consistent end-to-end architecture and offers compiler support for hardware changes with low overhead for both, (ii) does not require modifications to processor microarchitecture, so could be applied to different architectures without extension or CPU support, (iii) can be added to scalable systems such as servers and networks as it is included in the SoC, not the processor design, (iv) can authenticate not only the hardware but also the software source (v) does not directly affect cache and TLB performance in a system that has sufficient resources.

## VI. Conclusion

We introduced a fully end-to-end and comprehensive framework solution to both software obfuscation and trusted execution. Compared to existing trusted execution and software obfuscation architectures, our framework significantly improves comprehensiveness and generalizability. In the prototype, hardware overhead and compiler costs are much lower compared to existing architectures. Our architecture combines two technology-independent ideas: 1) While compiling, the program is encrypted with the PUF-based key of the target hardware and generates its signature. 2) Before execution on the target hardware, ERIC's hardware architecture decrypts the program with its PUF-based key and after verifying with the signature that the program is in the original version, executes it. While effective at improving both the safety of the program and authentication of executions on hardware, ERIC Framework is also configurable and simple to implement. Our future work will focus on improving the parallelism, performance, and scalability abilities of the architecture so that the framework can be implemented in distributed systems and implemented on servers. We also aim to bring RSA-based key generation and usage to ERIC.